\documentclass[aps,prl,reprint,groupedaddress,floatfix,superscriptaddress,amsmath]{revtex4-1}
\usepackage[latin9]{inputenc}
\makeatletter

\usepackage{changes}
\usepackage{graphicx}
\usepackage{color}
\usepackage{epsfig}
\usepackage{lineno}

\makeatother

\begin{document}
\title{Experimental identification of electric dipoles induced by magnetic monopoles in Tb$_{2}$Ti$_{2}$O$_{7}$}

\author{Feng\,Jin}
\affiliation{Beijing National Laboratory for Condensed Matter Physics, Institute of Physics, Chinese Academy of Sciences, Beijing 100190, China}
\affiliation{Department of Physics, Renmin University of China, Beijing 100872, China}

\author{Changle\,Liu}
\affiliation{State Key Laboratory of Surface Physics and Department of Physics, Fudan University, Shanghai 200433, China}

\author{Anmin\,Zhang}
\affiliation{School of Physical Science and Technology, Lanzhou University, Lanzhou 730000, China}

\author{Xiaoqun\,Wang}
\affiliation{Department of Physics and Astronomy, Shanghai Jiao Tong University, Shanghai 200240, China}

\author{Gang\,Chen}
\affiliation{State Key Laboratory of Surface Physics and Department of Physics, Fudan University, Shanghai 200433, China}

\author{Xuefeng\,Sun}
\affiliation{Hefei National Laboratory for Physical Sciences at Microscale, University of Science and Technology of China, Hefei, Anhui 230026, China}

\author{Qingming\,Zhang}
\email[e-mail:]{qmzhang@ruc.edu.cn}
\affiliation{School of Physical Science and Technology, Lanzhou University, Lanzhou 730000, China}
\affiliation{Beijing National Laboratory for Condensed Matter Physics, Institute of Physics, Chinese Academy of Sciences, Beijing 100190, China}

\begin{abstract}
The fundamental principles of electrodynamics allow an electron carrying both electric monopole (charge) and magnetic dipole (spin) but prohibit its magnetic counterpart. Recently it was predicted that the magnetic "monopoles"  carrying emergent magnetic charges in spin ice systems can induce electric dipoles. The inspiring prediction offers a novel way to study magnetic monopole excitations and magnetoelectric coupling. However, no clear example has been identified up to now. Here, we report the experimental evidence for electric dipoles induced by magnetic monopoles in spin frustrated Tb$_{2}$Ti$_{2}$O$_{7}$. The magnetic field applied to pyrochlore Tb$_{2}$Ti$_{2}$O$_{7}$ along {[}111{]} direction, brings out a "3-in-1-out" magnetic monopole configuration, and then induces a subtle structural phase transition at $H_{c} \sim$ 2.3 T. The transition is evidenced by the non-linear phonon splitting under magnetic fields and the anomalous crystal-field excitations of Tb$^{3+}$ ions. The observations consistently point to the displacement of the oxygen O$''$ anions along {[}111{]} axis which gives rise to the formation of electric dipoles. The finding demonstrates that the scenario of magnetic monopole having both magnetic charge and electric dipole is realized in Tb$_{2}$Ti$_{2}$O$_{7}$ and sheds light into the coupling between electricity and magnetism of magnetic monopoles in spin frustrated systems.
\end{abstract}

\maketitle

The interplay of electricity and magnetism is always a central topic in fundamental physics. In recent decades, the topic has received renewed attention in many exciting fields such as multiferroics, magnetoelectrics and spintronics~\cite{Fiebig, Cheong2007}. The coexistence of various degrees of freedom (lattice, charge and spin) and the entanglement between them, output a large variety of unusual effects and responses~\cite{Khomskii2009}. These systems represent excellent platforms for quantum control and engineering in both fundamental research and practical applications.

Recently, Khomskii \emph{et al.} proposed a fascinating scenario that the magnetic monopoles in the spin ice compounds can induce electric dipoles~\cite{Khomskii, Khomskii2010}. The strong coupling between the magnetic monopoles and the induced electric dipoles enables us to study and control such exotic magnetic monopoles by means of electric fields, and offers a novel way to develop potential applications in quantum computation. In solids, the magnetic monopoles are realized as emergent particles but not as elementary~\cite{Castelnovo2008}. And the most prominent examples of magnetic monopoles are topological defects of spin ice textures in pyrochlore systems, i.e., each tetrahedra with three spins pointing "in" and one spin pointing "out" (3-in-1-out, monopole) or vice versa (3-out-1-in, antimonopole)~\cite{Gingras2009, Gardner}. Beside spin ice textures, a giant magneto-electric coupling is also regarded as one of the essential ingredients to realize multiferroicity~\cite{Stojanovic, Dong}.

The effort of looking for the ideal candidate demonstrating magnetic monopole and electric dipoles and their coupling, is thus focused on the pyrochlore family with strong magneto-electric coupling. Very recently, it was theoretically pointed out that one of the family members, Tb$_{2}$Ti$_{2}$O$_{7}$, is perhaps a good candidate to observe electric dipoles induced by magnetic monopoles~\cite{Jaubert}. On the experimental side, to the best of our knowledge, the induced electric dipoles and their coupling with magnetic dipoles have not been clearly identified so far.

In the present article, we employed magneto-Raman scattering technique to search for electric dipoles induced by magnetic monopoles in the pyrochlore Tb$_{2}$Ti$_{2}$O$_{7}$. The alternating magnetic monopole/anti-monopole structure are stabilized by magnetic fields along {[}111{]} axis~\cite{Sazonov}. Meanwhile, the formation of electric dipoles are manifested by means of subtle structural changes that are captured by our high-precision Raman measurements under magnetic fields. The application of magnetic fields results in the unusual non-linear phonon splitting, the anomalous splitting of crystal field excitations (CFEs) and the emergence of new CFEs. The observations can be consistently and well explained in term of the shift of the oxygen (O$''$) along {[}111{]} axis that gives rise to the electric dipoles. The findings demonstrate that electric dipoles induced by magnetic monopoles and the strong coupling between them are unambiguously identified in Tb$_{2}$Ti$_{2}$O$_{7}$. This opens new possibilities to control magnetic monopoles with electric fields.

The Tb$_{2}$Ti$_{2}$O$_{7}$ single crystal was grown by the floating-zone technique and was cleaved before experiments. Confocal micro-Raman measurements were performed with a backscattering configuration using a Jobin Yvon T64000 system and a 532-nm diode-pumped solid-state laser. The laser power was kept at a level of 500 $\mu W$ to avoid overheating. Magnetic fields were generated up to 9 T using a superconducting magnet (Cryomagnetics), and the direction of magnetic field was along the {[}111{]} axis with an accuracy of $\pm$ 2$^{\circ}$.

The Raman spectra taken at 10 K and 0 T are shown in Fig.~1a, in which three strong phonon modes appear at $\sim$289 (F$_{2g}$), 320 (E$_{g}$) and 511 (A$_{1g}$) cm$^{-1}$. According to previous reports~\cite{Lummen, Chernyshev}, the F$_{2g}$ mode is assigned as the combined vibration of O$'$ (48f) and O$''$ (8a) anions while the A$_{1g}$ and E$_{g}$ modes are solely due to the vibration of O$'$ anions. With the application of magnetic field along [111] axis, the A$_{1g}$ mode is nearly unchanged but the F$_{2g}$ mode shows a clear splitting and eventually evolves into two well-resolved modes (P1 and P2) at 9 T (Fig.~1b).

The photoluminescence origin of P1 or P2 can be easily excluded since they remain unchanged under different excitation sources (Fig.~1a). The magnetic origin is also unlikely because of the very small magnetic exchange energy in Tb$_{2}$Ti$_{2}$O$_{7}$ ($\theta$ = -19 K)~\cite{Gardner2}. And the CFE origin is incompatible with the following characteristics of the two modes: (1) the P1 and P2 intensities are almost one order of magnitude larger than that of a typical CFE at $\sim$100 cm$^{-1}$~\cite{SuppleM}; (2) the P1 and P2 energies are well below that of the CFE ($\sim$339 cm$^{-1}$) revealed by neutron scattering experiments~\cite{Ruminy} and our Raman experiments~\cite{SuppleM}); and (3) the P2 energy, which goes to saturation with increasing fields, exhibits a field dependence distinguished from that of a typical CFE, which normally manifests a pronounced linear field dependence at high fields~\cite{SuppleM}.

\begin{figure}[t]
\centering
\includegraphics[width=8cm]{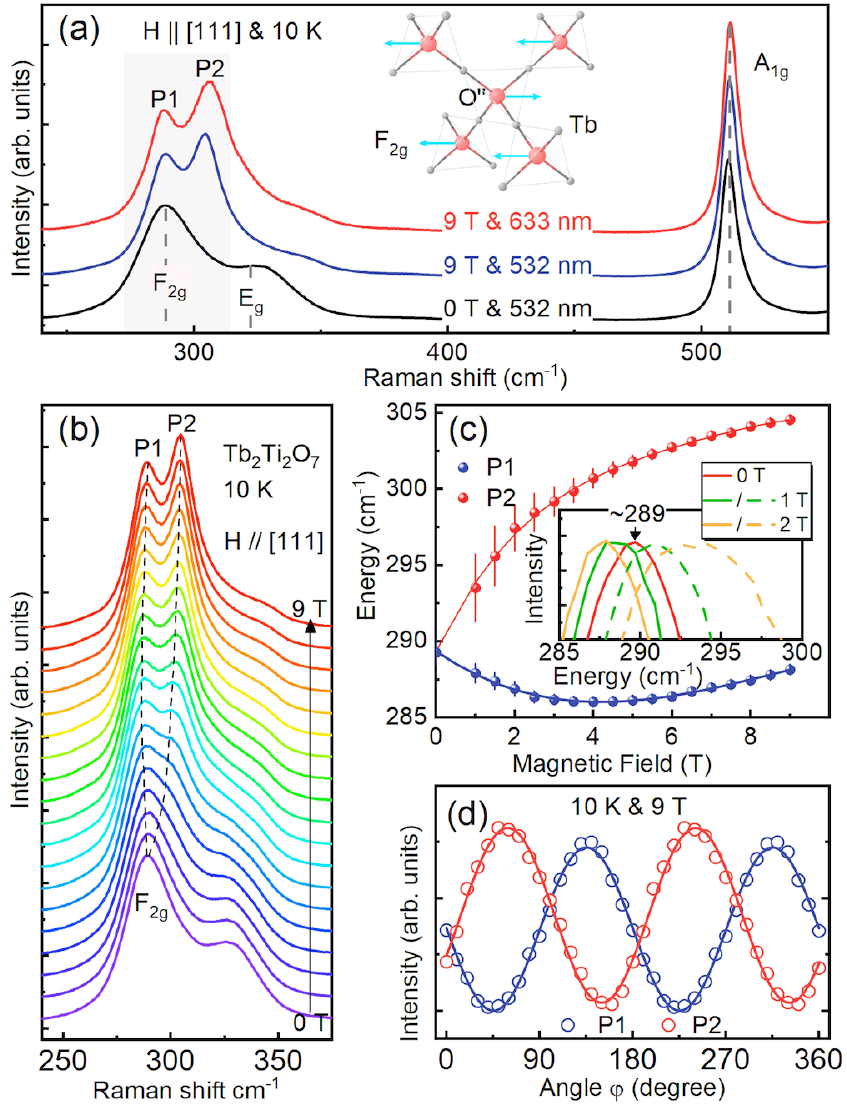}
\caption{Splitting of the F$_{2g}$ phonon mode with magnetic fields along {[}111{]} direction. (a), Raman spectra of Tb$_{2}$Ti$_{2}$O$_{7}$ taken at 0 and 9 T with different excitation sources. Inset: vibrational pattern of the  F$_{2g}$ mode in which the O$'$ ions are omitted for clarification. (b), Magnetic field evolution of the F$_{2g}$ mode collected with e$_{s}\perp$ e$_{i}$ $\parallel [11\bar{2}]$. (c), Field dependence of the energies of P1 and P2 modes. Inset: the splitting of F$_{2g}$ mode at low fields, collected with two polarization configurations (solid and dashed curve). (d), Polarization dependence of the intensities of P1 and P2 modes at 9 T. The raw spectra can be found in~\cite{SuppleM}.}
\end{figure}

\begin{figure*}[t]
\centering 
\includegraphics[width=15cm]{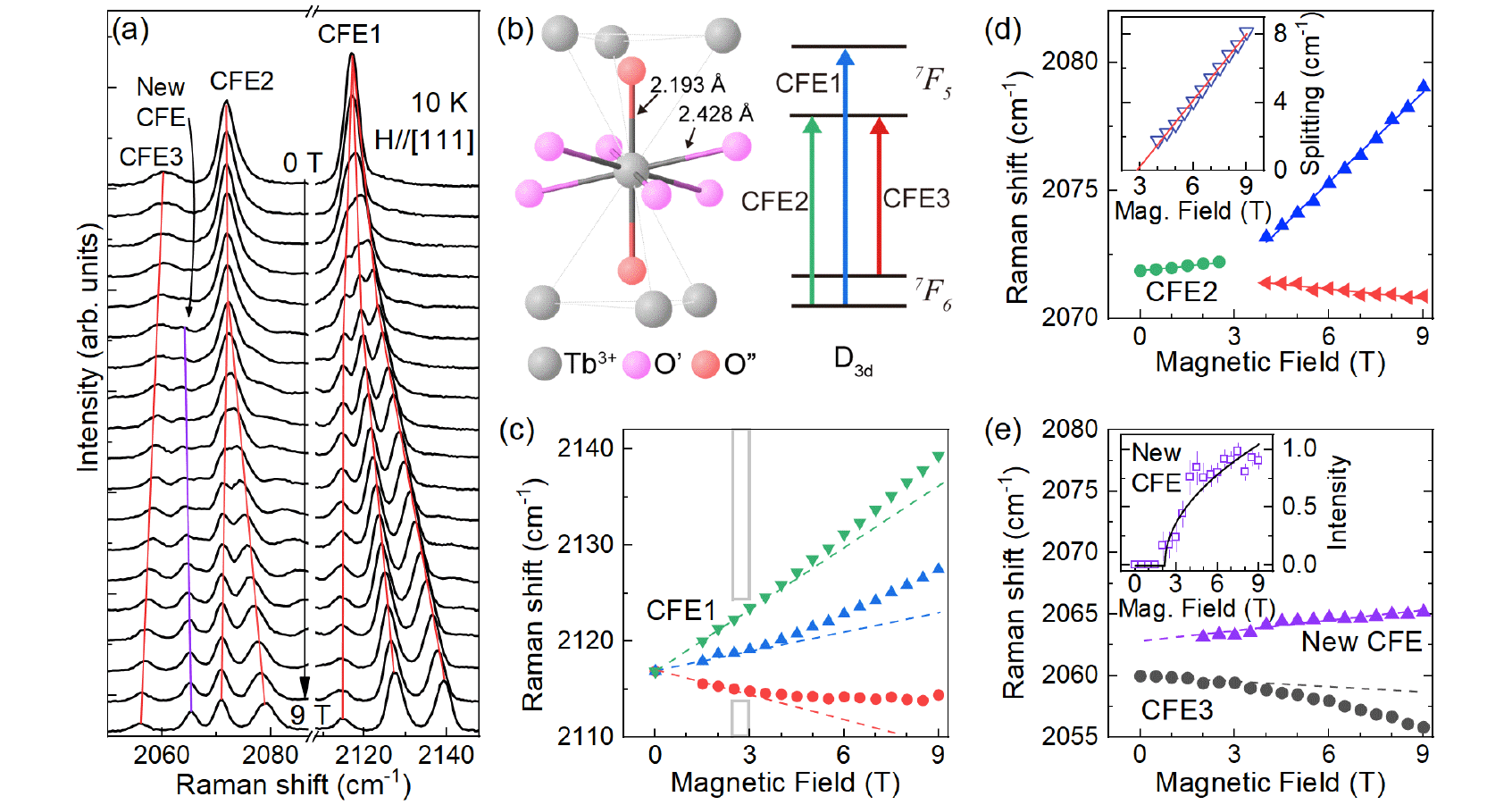}
\caption{Distortion of the local geometry around the Tb$^{3+}$ ions under magnetic fields. (a), Field evolution of the CFE spectra between 2050 and 2150 cm$^{-1}$. (b), Local geometry around the Tb$^{3+}$ ions along with the selected crystal-field energy levels and excitations. (c), Zeeman splitting at low fields and the non-linear splitting at higher fields of the CFE1. The dashed curves are linear fits to the data below 3 T. (d), Field dependence of the CFE2 energies, exhibits a further splitting at $\sim$3 T (inset, Energy$_{blue}$-Energy$_{red}$). \textbf{e}, Field dependence of the energies of the CFE3 and the new CFE. The inset shows the integrated intensities of the new CFE. Solid curve is the fit to $I_0 + I \sqrt{H- H_c}$.}
\end{figure*}

By ruling out the above origins, we can safely and simply attribute the P1 and P2 modes to the splitting of the F$_{2g}$ phonon mode. This is strongly supported by the field dependence of their energies (Fig.~1c) and the polarization dependence of their intensities (Fig.~1d). At $H$ = 9 T, the P2 mode locates at $\sim$16 cm$^{-1}$ above the P1 mode. The energy difference of the two modes decreases with decreasing fields and the two modes eventually merge into a single mode at $\sim$0 T. Although it is not easy to precisely extract the P1 and P2 positions at low magnetic fields through fitting process, the polarized Raman spectra (inset of Fig.~1c) clearly show that the energy difference between the two modes approaches to zero with decreasing fields, suggesting that the P1 and P2 modes stem from the split of the F$_{2g}$ phonon. Moreover, the polarization dependence of the P1 and P2 intensities exhibits a clear anti-phase correlation (Fig.~1d), indicating they share the same origin, i.e., the F$_{2g}$ splitting. Then we conclude that the F$_{2g}$ mode splits into two modes with the application of magnetic field along [111] axis and the splitting suggests either a lattice modulation or a redistribution of electrons, or both.

To further explore the field-induced effects in Tb$_{2}$Ti$_{2}$O$_{7}$, let's turn to the CFEs which directly probe the local environments around Tb$^{3+}$ ions. Generally, magnetic fields split the CFEs and their energies are expected to linearly depend on fields due to Zeeman effects. However, if there exists a strong field-induced lattice modulation that substantially affects the crystal field environments, the CFEs will behave anomalously, such as the nonlinear field dependence of the CFE energies, further splitting of CFEs and the emergence of new CFEs at non-zero field.

The anomalous evolutions of several CFEs under H $\parallel$ {[}111{]} are illustrated in Fig.~2a. At zero field, three CFEs are observed at $\sim$2060 (CFE3), 2072 (CFE2) and 2117 (CFE1) cm$^{-1}$. According to the crystal field calculations~\cite{Klekovkina}, the three CFEs involve the transitions from the $^{7}$F$_{6}$ manifolds to the $^{7}$F$_{5}$ manifolds, as schematically shown in Fig.~2b. The CFE1 splits into three strong peaks with increasing magnetic fields. The peak energies (Fig.~2c) exhibit nearly a linear field dependence below $\sim$2.5 T due to the Zeeman effect, and clearly deviate from the linear behaviors at H $>$ 2.5 T. The non-linear field behaviors at H $>$ 2.5 T are also witnessed by the CFE3 (Fig.~2e) and many other CFEs~\cite{SuppleM}, systematically suggesting a change of crystal field environments of Tb$^{3+}$ ions.

The change of crystal field environments is more clearly evidenced by the splitting of the CFE2 and the emergence of a new CFE at $\sim$2.5 T (Fig.~2d and 2e), which point to a distortion of the local geometry around Tb$^{3+}$ ions. Figure 2d shows the field dependence of the CFE2 energies, which is linear at low fields and then splits into two peaks between 2-3 T. The CFE2 splitting starting at $\sim$2.5 T but not 0 T (inset of Fig.~2d), is quite unusual and needs to be understood with the distortion of the local geometry around Tb$^{3+}$ ions beyond the simple Zeeman effect. Meanwhile, a new CFE accompanying the CFE2 splitting appears between 2060 and 2070 cm$^{-1}$ (Fig.~2a). The normalized integrated intensities (I.I.) of the new CFE are shown in the inset of Fig.~2e and a transition-like upturn is seen above a critical field $H_c =$ 2.3 T. This demonstrates that the new CFE does not originate from the Zeeman splitting of any CFE but is related to the distortion of the local geometry around Tb$^{3+}$ ions.

The above findings, including the F$_{2g}$ splitting and the anomalous behaviors of CFEs under magnetic fields, allow us to conclude that a field-induced subtle structural transition occurs at $H_c$ in Tb$_{2}$Ti$_{2}$O$_{7}$. Now the question is how to understand the structural transition. In frustrated spin systems, the lattice structure strongly depends on their magnetic structure~\cite{Stojanovic, Kimura, Walker}. For Tb$_{2}$Ti$_{2}$O$_{7}$, with increasing magnetic fields along {[}111{]} axis, it undergoes a magnetic transition from the zero field spin ice/liquid state to the magnetic monopole structure (3-in-1-out/3-out-1-in, Fig.~3a) at a very small filed $H_0$~\cite{Sazonov}. This hints that the field-induced structural transition observed in Tb$_{2}$Ti$_{2}$O$_{7}$ stems from the field-induced magnetic monopole structure.

Several microscopic approaches, such as inverse Dzyaloshinskii-Moriya (DM) interaction~\cite{Katsura,Sergienko1}, spin-dependent $p$-$d$ hybridization~\cite{Jia} and exchange striction~\cite{Sergienko2}, have been proposed to understand the magnetically driven lattice modulation. The hybridization and/or exchange based scenarios seem unlikely here, since $4f$ electrons are significantly screened by the outer $5s$ and $5p$ orbitals and little participate in bonding. Actually, the field-induced structural transition observed here can be naturally understood in term of magnetic monopole configuration and the strong spin-orbit coupling in rare-earth ions.

\begin{figure}[t]
\centering 
\includegraphics[width=8.5cm]{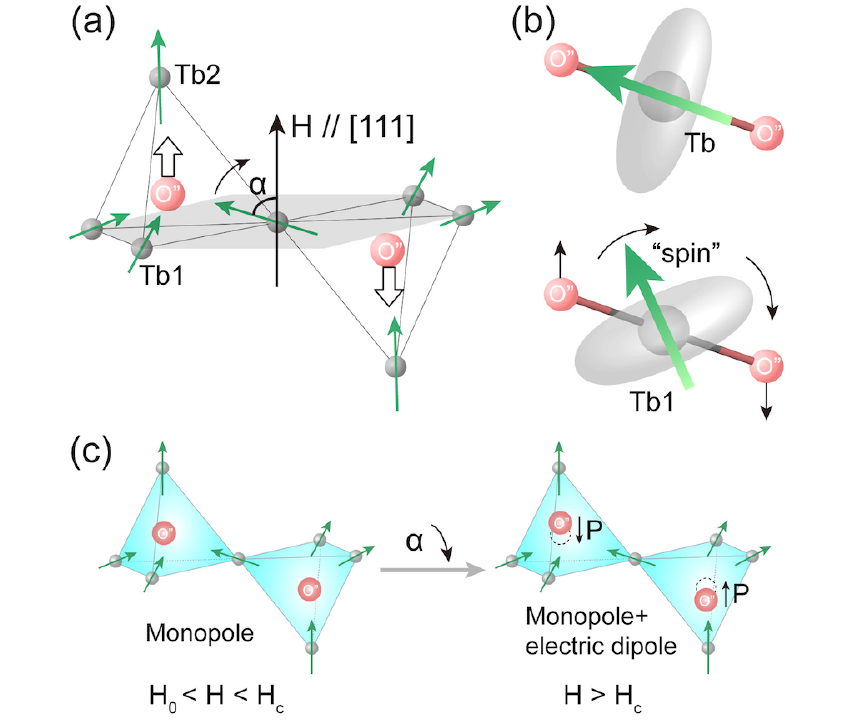}
 \caption{Schematic illustration of magnetoelastic mechanism. (a), Electric dipoles and magnetic monopoles induced by {[}111{]}-magnetic field. Note that the displacement of O$''$ ions is along the {[}111{]} direction and away from the Kagom$\acute{e}$ plane (shaded plane). $\alpha$ is the angle between the Tb1 magnetic moments and the applied field. (b), Schematic of the oblate $4f$ charge cloud of Tb1$^{3+}$ ions (Upper) and its rotation under magnetic fields (Lower). (c), Field induced magnetic monopole phase ($H_0 < H < H_c$), monopole and electric dipole coupled phase ($H > H_c$).}
\end{figure}

\begin{figure}[t]
\centering 
\includegraphics[width=8cm]{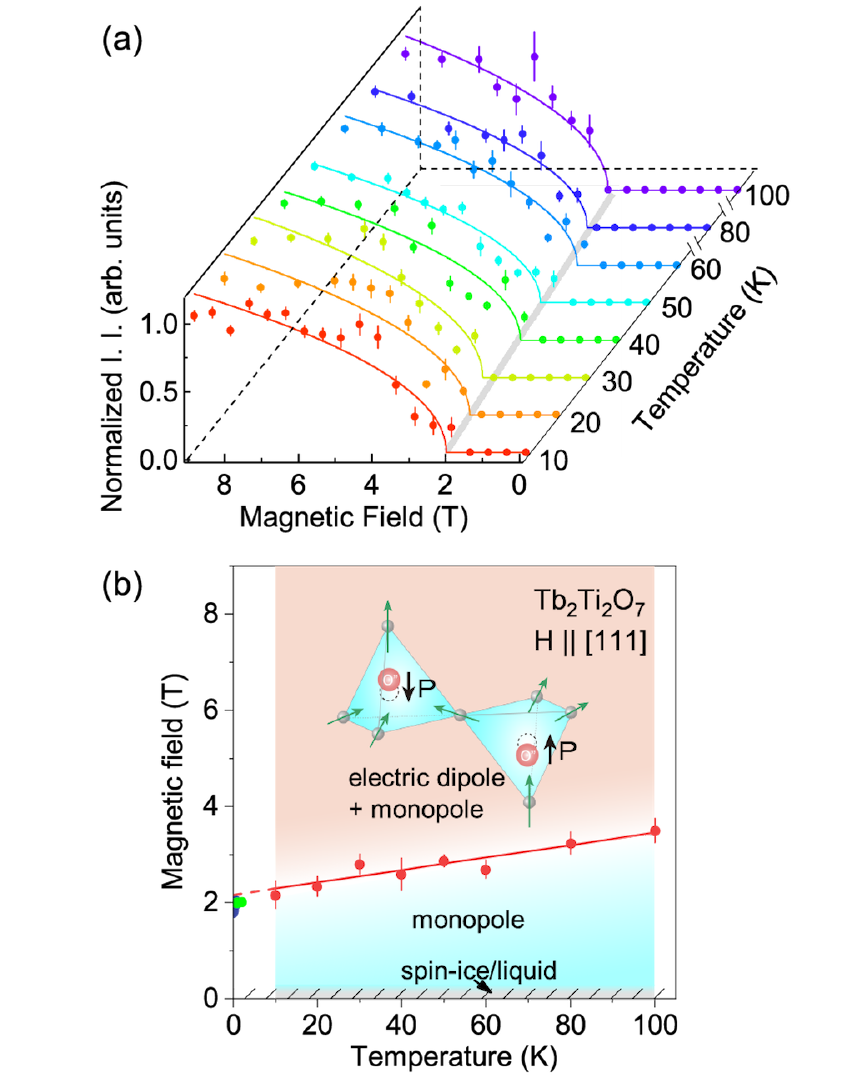}
 \caption{Magnetic field - temperature phase diagram of Tb$_{2}$Ti$_{2}$O$_{7}$ for the {[}111{]}-magnetic field. (a), Field-dependence of I. I. of the new CFE normalized to the value at 9 T as a function of temperature. The colored solid curves are fits to $I_0 + I \sqrt{H- H_c}$. The gray curve traces the evolution of $H_c$. The standard errors (SE) of the normalized I. I. are taken from the Lorentzian fitting. (b), Magnetic field-temperature phase diagram. The error bars correspond to the SEs of $H_c$ in the fitting for (a). The solid blue and green circles are the $H_s$ values taken from Ref.~\cite{Legl} and Ref.~\cite{Hirschberger}, respectively.}
\end{figure}

Unlike 3d ions, rare-earth ions possess a very strong spin-orbit coupling ($\sim 1eV$), which tightly locks the spin and the orbital angular momentum together, where the orbital angular momentum results from the spatially anisotropic $4f$ wave functions that can be simplicity envisioned as oblate electron charge cloud (equatorially expanded, upper of Fig.~3b)~\cite{Engdahl}. Due to the strong spin-orbit coupling, the orientation of the anisotropic shaped electron charge cloud is rigidly attached to the direction of the spin moment. At zero field, the system stays in the spin ice/liquid phase with magnetic moments randomly pointing into or out of tetrahedra. Meanwhile, the O$''$ ions reside at the center of Tb$_{4}$ tetrahedra because of the equivalence of the four Tb$^{3+}$ ions in a tetrahedron. The magnetic field along {[}111{]} axis breaks the equivalence of the Tb$^{3+}$ ions between the ones in the Kagom$\acute{e}$ planes (Tb1) and that out of Kagom$\acute{e}$ planes (Tb2) (see Fig.~3a). Upon increasing field, the magnetic moment of Tb2 stays parallel to the field while that of Tb1 receives a moment tilting from its original direction towards the field direction (see lower of Fig.~3b)~\cite{Sazonov}. Meanwhile, the strong spin-orbit coupling will drive the charge cloud ($4f$ orbit) of Tb1 ions to rotate accordingly. This will increase the Coulomb repulsion between the charge clouds of Tb1 and O''. To minimize the overall energy, the O$''$ ions tend to displace away from the Kagom$\acute{e}$ planes along the {[}111{]} direction, see Fig.~3a. As a result, the induced electric dipoles emerge and the system develops an antiferroelectric order.

The scenario well explains our experimental observations. For H $<H_{c}$, the Coulomb repulsion is too small to drive the displacement of O$''$ ions. Therefore, the field dependence of the CFEs is governed by the Zeeman effect. On the other hand, the F$_{2g}$ mode is very sensitive to the breaking of the equivalence of O$''$-Tb1 and O$''$-Tb2 bonds, and splits into two branches even at H $< H_c$. For H $>H_{c}$, the O$''$-Tb1 and O$''$-Tb2 bonds are not further broken by the displacement of O$''$ ions. It explains why the P1 and P2 modes show only a kink in energy rather than a jump when crossing the transition. However, the real displacement of O$''$ ions changes the local geometry of Tb$^{3+}$ ions, resulting in the observed anomalous behaviors of CFEs. The peak bifurcation of CFE2 above $H_{c}$ is a natural consequence of the inequivalence of crystal field environments between Tb1 and Tb2. And the variations of CF wave functions caused by the displacement in principle relax the CF transition rules and render some transitions visible in Raman channel, i.e., the emergence of the new CFE peak. Then we conclude that an antiferroelectric transition occurs in Tb$_{2}$Ti$_{2}$O$_{7}$ at $H_c =$ 2.3 T.

To track the temperature dependence of the antiferroelectric transition, we further carried out field-dependent Raman measurements under various temperatures. Figure 4a displays the field dependence of the normalized I. I. of the new CFE at these temperatures, where $H_c$ are extracted by fitting with the order-parameter-like function I.I. $\propto \sqrt{H - H_c}$. This gives us the magnetic field-temperature phase diagram (Fig.~4b). From 10 to 100 K, $H_c$ linearly  increases from $\sim$2.3 T to 3.5 T (see Fig.~4b). Interestingly, the extrapolated field at zero temperature ($\sim$2.1 T) agrees well with the field $H_2$ given by susceptibility~\cite{Legl} and $H_s$ by thermal conductivity~\cite{Hirschberger}. It suggests that the transition at $H_2$/$H_s$ reported by thermodynamic measurements, is related to the antiferroelectric transition observed here. The finding may be a key to understand the anomalous thermal conductivity observed in Tb$_{2}$Ti$_{2}$O$_{7}$~\cite{Hirschberger}.

Let's conclude the paper by discussing the key consequence of the field induced antiferroelectric transition in Tb$_{2}$Ti$_{2}$O$_{7}$, the emergence of electric dipole on its magnetic monopole. Tb$_{2}$Ti$_{2}$O$_{7}$ is thus identified here as the first example in which the magnetic "monopoles" have both magnetic charges and coupled electric dipoles. The study makes the close analogy of electricity and magnetism go even further than usually assumed, i.e., the counterpart of a point charge (electron) not allowed in the fundamental level, can be realized as an emergent particle in condensed matter systems. This may bring many new and intriguing possibilities and greatly extend the study of pyrochlore spin systems. For example, the coupling between magnetic monopoles and electric dipoles allow to study and control the monopoles by external electric fields, i.e., creation, elimination and separation of monopoles and antimonopoles.

\begin{acknowledgments}
This work was supported by the Ministry of Science and Technology of China (2017YFA0302904 \& 2016YFA0300504) and the NSF of China (11774419 \& 11474357).

F.J. and C.-L.L. contributed equally to this work.
\end{acknowledgments}

\end{document}